\documentclass[ reprint,
nofootinbib,
 amsmath,amssymb,
 aps,
 superscriptaddress
]{revtex4-2}
\usepackage{xcolor}
\usepackage{graphicx}
\usepackage{dcolumn}
\usepackage{bm}
\usepackage{multirow}

\usepackage{hyperref}
\newcommand{\be}{\begin{equation}}
\newcommand{\ee}{\end{equation}}
\newcommand{\beq}{\begin{equation}}
\newcommand{\eeq}{\end{equation}}
\newcommand{\ba}{\begin{aligned}}
\newcommand{\ea}{\end{aligned}}

\newcommand{\bea}{\begin{eqnarray}}
\newcommand{\eea}{\end{eqnarray}}

\newcommand{\cN}{\mathcal{N}}

\newcommand{\cM}{\mathcal M}

\newcommand\bi{\begin{itemize}}
\newcommand\ei{\end{itemize}}

\def\unit{{1\kern-.65ex {\rm l}}}
\def\1{{1\kern-.65ex {\rm l}}}

\begin{document}


\begin{flushright}
{\tt\normalsize ZMP-HH-25/18}\\
\end{flushright}

\title{K-Points and Type IIB/Heterotic Duality with NS5-Branes}

\author{Jeroen Monnee}
\affiliation{%
 II. Institut f\"ur Theoretische Physik, Universit\"at Hamburg,\\ Notkestrasse 9, 22607 Hamburg, Germany
}%
\author{Timo Weigand}
\affiliation{%
 II. Institut f\"ur Theoretische Physik, Universit\"at Hamburg,\\ Notkestrasse 9, 22607 Hamburg, Germany
}%
\affiliation{%
 Zentrum f\"ur Mathematische Physik, Universit\"at Hamburg\\
 Bundesstra\ss e 55, 20146 Hamburg, Germany
 }
\author{Max Wiesner}%
\affiliation{%
 II. Institut f\"ur Theoretische Physik, Universit\"at Hamburg,\\ Notkestrasse 9, 22607 Hamburg, Germany
}%

\date{\today}

\begin{abstract}
We revisit type II$_0$ limits in the vector multiplet moduli space of compactifications of Type IIB string theory on Calabi--Yau threefolds. These limits are special because they cannot be described using mirror symmetry. While we showed in previous work that the gravitational duality sector emerging in such limits corresponds to a weakly coupled heterotic string compactified on K3 times a string-sized torus, our focus here is on the additional field theory sectors that decouple from gravity in these asymptotic regimes. Concretely, we argue that, in the dual heterotic description, these field theory sectors correspond to spacetime-filling NS5-branes wrapping the heterotic torus. The existence of such non-perturbative field theory sectors manifests itself in an exponential dependence of the prepotential on the heterotic dilaton at leading order. Applied to K-point limits in one-parameter moduli spaces, our results imply that these limits are in perfect agreement with the Distance and Emergent String Conjectures. More precisely, the light tower of states predicted by the Distance Conjecture arises from excitations of a weakly coupled heterotic string compactified to four dimensions on a background containing spacetime-filling NS5-branes.
\end{abstract}

\maketitle

\section{Introduction}

An increasing amount of evidence suggests that consistent quantum gravity theories exhibit universal behaviour near asymptotic boundaries of their moduli space. 
 The Swampland Distance Conjecture \cite{Ooguri:2006in} formulates this in terms of a tower of infinitely many weakly coupled states which are conjectured to become light exponentially fast towards regions in moduli space that lie at infinite distance. According to the Emergent String Conjecture \cite{Lee:2019oct}, these towers can always be interpreted, at least for theories reducing to Einstein gravity at low energies, as the Kaluza--Klein towers associated with a decompactifying dimension or as excitations of a weakly coupled string, even if the theory in the interior of the moduli space need not admit a description as a string theory. 
 Both circles of ideas have been tested with great success in the framework of string and M-theory compactifications \cite{Lee:2018urn,Lee:2019oct,Lee:2019apr,Rudelius:2023odg,Lee:2021qkx,Lee:2021usk,Chen:2024cvc,Alvarez-Garcia:2023gdd,Alvarez-Garcia:2023qqj,Alvarez-Garcia:2021pxo,Marchesano:2019ifh,Baume:2019sry,Xu:2020nlh,Lee:2019tst,Klaewer:2020lfg,Basile:2022zee,Etheredge:2023odp,Aoufia:2024awo,Calderon-Infante:2024oed,Friedrich:2025gvs,Gkountoumis:2025btc,Monnee:2025ynn}; bottom-up arguments for the Distance Conjecture and the Emergent String Conjecture from several angles have been developed in \cite{Basile:2023blg,Basile:2024dqq,Herraez:2024kux,Cribiori:2023ffn,Bedroya:2024ubj,Kaufmann:2024gqo}, respectively. 
 Consequences of the Emergent String Conjecture for the classification of Quantum Gravity theories in various dimensions have been analysed in \cite{Kim:2024eoa,Etheredge:2025ahf,Grieco:2025bjy}.

 Within the complex structure moduli space of Type IIB compactifications on Calabi--Yau threefolds, the Emergent String Conjecture follows by
  combining insights from algebraic \cite{schmid,CKS,Grimm:2018ohb,Grimm:2018cpv,Grimm:2019bey,Grimm:2021ikg,Palti:2021ubp,Bastian:2023shf,vandeHeisteeg:2022gsp,Monnee:2024gsq} and geometric \cite{Friedrich:2025gvs,Monnee:2025ynn} mixed Hodge structures.
   In \cite{Friedrich:2025gvs} it has been shown that near limits of type II, the Type IIB string is dual to either a critical heterotic or another Type II string, depending on whether the 3-fold components into which the Calabi--Yau degenerates intersect over K3 surfaces or Abelian surfaces. 
 Special examples of such type II degenerations  are the so-called K-points \cite{vanstraten2017calabiyauoperators,Joshi:2019nzi}, which arise in complex structure moduli spaces of dimension one. What makes them somewhat mysterious at first sight is that they cannot be connected to a large complex structure locus and hence do not admit a straightforward geometric Type IIA mirror dual description \cite{doran2016mirrorsymmetrytyurindegenerations,Grimm:2018cpv, Grimm:2018ohb}. 
 Nonetheless, according to the analysis of \cite{Friedrich:2025gvs}, they describe 
 a weakly coupled heterotic string on K3 $\times T^2$ at special values for the torus complex structure. In this short note, we provide further evidence for this interpretation by identifying the origin of non-perturbative field theory sectors decoupling from gravity in these limits. This is particularly important for the interpretation of the prepotential near K-points \cite{vanstraten2017calabiyauoperators,Joshi:2019nzi,Bastian:2023shf}, whose special properties have been stressed recently \cite{Hattab:2025aok}.

 Key to the analysis is the distinction between  the gravitational and the field theory sector of the asymptotic theory. 
 As scrutinized in a series of recent works \cite{Marchesano:2023thx,Marchesano:2024tod,Castellano:2024gwi,Blanco:2025qom} (see also \cite{Cota:2022yjw,Cota:2022maf}),
 along generic infinite distance loci
 a weakly coupled gravitational sector and a strongly coupled field theory sector, decoupled from gravity, can co-exist. For emergent string limits, this means that while the gravitational duality frame is set by a perturbative, weakly coupled critical string, there can nonetheless appear non-perturbative field theoretic sectors that decouple from the gravitational dynamics. In particular, the secondary singularity type describing the limit in the language of asymptotic Hodge theory \cite{Grimm:2018cpv,Grimm:2018ohb} contains information about the rank of the field theory sector coupled to gravity \cite{Marchesano:2023thx,Marchesano:2024tod,Castellano:2024gwi,Monnee:2025ynn}.

 As we will illustrate explicitly in the context of a 2-parameter model
 extensively studied already in \cite{Candelas:1993dm,Kachru:1995fv}, the presence of a non-perturbative field theory sector is reflected 
  in the leading order behaviour of the prepotential:
  It contains a term exponential in the local flat coordinate that goes to infinity in the type II degeneration, 
 as opposed to a purely polynomial leading order dependence (with subleading exponential corrections).
 In the heterotic  duality frame on K3 $\times T^2$, we interpret the non-perturbative subsector as the gauge sector from a spacetime-filling wrapped NS5-brane, and its charged BPS states as wrapping modes of an E-string  contained in the NS5-brane. Nonetheless, the gravitational duality frame is set by the perturbative heterotic string, which is weakly coupled. If the non-perturbative sector is completely Higgsed, the theory reduces to a weakly coupled gravity sector coupled to a hypermultiplet originating from the non-perturbative gauge sector. This is the origin of the non-perturbative terms in the prepotential which in particular govern the K-point; the latter fact has been stressed recently in \cite{Hattab:2025aok} in the context of applying the Emergence Proposal \cite{Heidenreich:2017sim,Grimm:2018ohb,Heidenreich:2018kpg,Palti:2019pca} to K-points. For studies of the Emergence Proposal in other regimes of the vector multiplet moduli space of Calabi--Yau compactifications, see~\cite{Blumenhagen:2023yws,Blumenhagen:2023tev,Hattab:2023moj,Hattab:2024chf,Artime:2025egu,Blumenhagen:2025zgf}. 
 
 The main point of our short analysis is that the presence of non-perturbative terms in the prepotential does not imply that the gravitational duality frame is away from weak coupling. 
  Rather, we argue that K-points share the property of more general type II degenerations that the gravitational sector is weakly coupled and described by a critical string theory. The latter is perturbative up to the string scale of the dual emergent string, even though it can contain decoupling non-perturbative field theory sectors arising from NS5-branes in the dual heterotic language. The existence of such decoupled field theory sectors below the string scale is not in tension with the Emergent String Conjecture. In particular, the tower of states predicted by the Distance Conjecture arises as modes of the weakly coupled string. 
 
This note is organised as follows:
 In section \ref{sec_2parameter} we introduce a two-parameter model and its type II degeneration loci in complex structure moduli space. In sections \ref{sec_Delta1} and \ref{sec_Delta0} we investigate the physics of the type II$_1$ and, respectively, 
type II$_0$ degenerations of this model, each time beginning with the Type IIB interpretation and then matching the findings to the perturbative heterotic dual description. In Section \ref{sec_DeltatoK} we relate the type II$_0$ locus to a K-point by a Higgsing process. Our general conclusions are contained in Section \ref{sec_disc}.

\section{Type II Degeneration in a Two-Moduli Example}  \label{sec_2parameter}
Consider the Calabi--Yau threefold $X$ given by the degree-12 hypersurface in $\mathbb{P}_{1,1,2,2,6}$ first studied in detail in~\cite{Candelas:1993dm}.  As shown in~\cite{Kachru:1995fv}, compactifying Type IIA string theory on this manifold is dual (in the large volume regime) to the heterotic string on K3$\times T^2$. Here, we are interested in the mirror dual, i.e., Type IIB string theory compactified on the mirror manifold $Y$. The latter is given by the locus 
\begin{equation*}
    P= z_1^{12}+z_2^{12} +z_3^6+z_4^6+z_5^2 -12\psi z_1z_2z_3z_4z_5 -2\phi z_1^6z_2^6 =0\,,
\end{equation*}
for which $G=\mathbb{Z}_6^2 \times\mathbb{Z}_2$ acting on the $z_i$ is modded out and all orbifold singularities are resolved. The complex structure moduli space, $\cM_{\rm c.s.}$, of $Y$ is parametrized by $(\phi,\psi)$ and  thus has complex dimension two. Along special divisors $\Delta_i\subset \cM_{\rm c.s.}$ the threefold $Y$ becomes singular. Of particular interest for us is the divisor corresponding to $y\equiv\phi^{-2}=0$. Along this divisor, $Y$ degenerates 
into the union of two Fano threefolds $V_{1,2}$ intersecting over a K3-surface $Z$,
\begin{equation}\label{Ydegeneration}
    Y\to Y_{y=0} = V_1 \cup_Z V_2 \,.
\end{equation}
The properties of $Z$ depend on the value of $\psi$. In particular, the divisor $y=0$ has a point of tangency with the discriminant divisor 
\begin{equation}
    \Delta_C = (1-x)^2 -x^2y=0 \,,\qquad x=-\frac{\phi}{864\psi^6}\,,
\end{equation}
at $x=1$. We can obtain normal crossing divisors by resolving this point of tangency corresponding to a certain scaling limit for $\psi\to \infty$. To this end, we introduce the coordinates $(\epsilon, u)$ defined implicitly via~\cite{Kachru:1995fv}
\begin{equation}
    \phi=-\frac1\epsilon\,,\quad \psi=-\frac{1}{\sqrt{3}}\left(2^5\epsilon (1+\epsilon u)\right)^{-1/6}\,. 
\end{equation}
We can now define two divisors, $\Delta_0=\{x_0=0\}$ and $\Delta_1=\{x_1=0\}$, where 
\begin{equation}
    x_0 = \epsilon u\;\,,\quad x_1 = u^{-2}\,. 
\end{equation}
Notice that to reach a generic point along $\Delta_1$, we must scale $u\sim \epsilon^{-1}\to \infty$. This means that a generic point on either divisor corresponds to $\epsilon\to 0$ such that $Y$ degenerates as in~\eqref{Ydegeneration}, and we identify $\Delta_0$ and $\Delta_1$ as type II divisors in the nomenclature of~\cite{Grimm:2018ohb}. Furthermore, the degeneration is of Tyurin type  and therefore the analysis of~\cite{Friedrich:2025gvs} is directly applicable to these degenerations. The divisors differ in the secondary singularity type, with $\Delta_0$ being of type II$_0$ and $\Delta_1$ of type II$_1$.  
 To see this, recall that in the vicinity of a general singular locus $\Delta$, the limiting mixed Hodge structure on $H^3(Y,\mathbb{C})$ can be characterised by the graded spaces $\text{Gr}_\ell(\Delta)$, $\ell=0,\dots,6$. For type II$_b$ singularities these graded spaces are non-empty only for $\ell=2,3,4$, and the secondary singularity type $b$ is defined via ${\rm dim}({\rm Gr}_2) = {\rm dim}({\rm Gr}_4) = 2 +b$ \cite{Grimm:2018cpv,Grimm:2018ohb}.
 Furthermore, only the elements of ${\rm Gr}_4$ transform non-trivially under the monodromy around the singularity divisor in moduli space.
To determine $b$, we can therefore consider the period vector in the vicinity of $\Delta_0\cap \Delta_1$. An integral basis is given, to leading order, by~\cite{Lee:2019oct,Bastian:2021eom} 
\begin{equation}\label{eq:periods}
    \begin{pmatrix}
        X^0\\ X^1\\X^2\\F_2\\F_1\\F_0
    \end{pmatrix} = \begin{pmatrix}
       \frac{1}{\pi} (1 +\beta ^2 x_0) \vspace{6pt}\\\frac{i}{\pi}( 1 -\beta^2x_0)\vspace{6pt}\\ -\frac{\beta}{\pi}\sqrt{x_0}\vspace{6pt}\\\frac{\beta}{2\pi^2 i}\log(x_1)\sqrt{x_0} +\frac{2\beta}{i\pi^2}\sqrt{x_0} \vspace{6pt}\\ \frac{\log(x_0^2x_1) -\beta^2\left(\log(x_0^2x_1) -4\right)x_0}{8\pi^2}\vspace{6pt}\\ \frac{\log(x_0^2x_1) +\beta^2\left(\log(x_0^2x_1) -4\right)x_0}{8\pi^2 i}
    \end{pmatrix}\,,
\end{equation} 
 with
${\rm log}(x_0)$ and 
${\rm log}(x_1)$ flat local coordinates on the moduli space and $\beta=\frac{\Gamma(3/4)^4}{\sqrt{3}\pi^2}$.
As it turns out, the computation of the prepotential requires the expansion of the periods to higher order in $x_1$, which we provide in Appendix \ref{app-periods}.

\begin{figure*}
    \centering
    \includegraphics[width=0.79\linewidth]{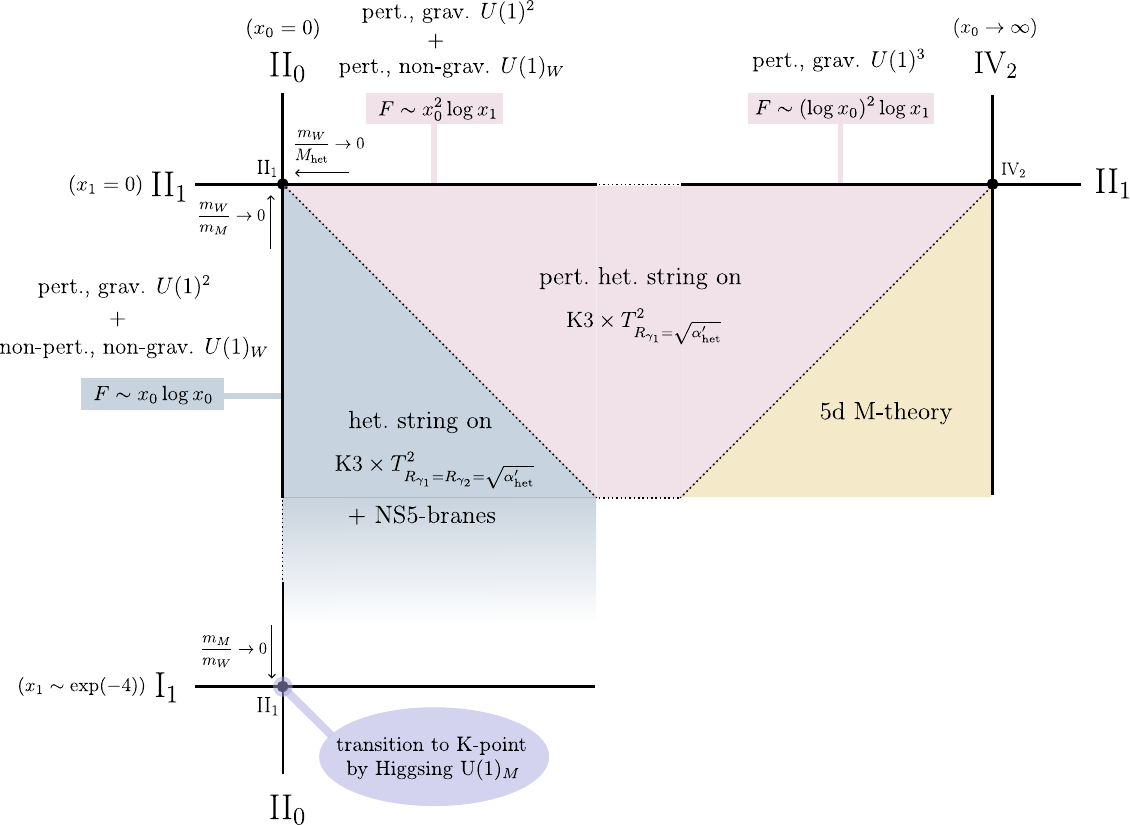}
    \caption{An overview of the different duality frames that emerge along the various singular divisors discussed in this work. In addition, we have included the local expansions for the prepotential $F$, as well as the number of perturbative / non-perturbative and gravitationally decoupled electric $U(1)$s in each regime. The non-gravitational $U(1)$ is part of a Seiberg--Witten like subsector with an associated W-boson of mass $m_W$ and a monopole of mass $m_M$. We have also highlighted in purple the intersection of the $\mathrm{II}_0$ and conifold ($\mathrm{I}_1$) divisors, where the transition to the K-point is performed by Higgsing the decoupled magnetic  U(1)$_M$.}
    \label{fig:duality-frames}
\end{figure*}
 
Consider now the monodromy around $x_1=0$ corresponding to the transformation of the period vector under $x_1 \mapsto e^{2\pi i}x_1$. Given that three of the six periods are logarithmic in $x_1$, the associated elements in $H^3(Y)$ transform non-trivially under the monodromy. 
 We hence conclude that $\dim({\rm Gr}_{\ell=2,4}(\Delta_1))=3$ and $\dim({\rm Gr}_{3}(\Delta_1))=0$ such that, indeed, $\Delta_1$ is a type II$_1$ divisor. Similarly, for the monodromy around $x_0=0$ induced by $x_0\mapsto e^{2\pi i} x_0$, only the two elements of $H^3(Y)$ for which the periods are logarithmic in $x_0$ pick up additional charges. Therefore
$\dim({\rm Gr}_{\ell=2,4}(\Delta_0))=2$ and $\dim({\rm Gr}_{3}(\Delta_0))=2$, which identifies $\Delta_0$ as a type II$_0$ divisor. 

Figure \ref{fig:duality-frames} provides a sketch of the location of the divisors $\Delta_0$ and $\Delta_1$ in the complex structure moduli space of $Y$ and indicates the duality frames emerging in the vicinity of these divisors as we discuss in detail in the following.

\section{Light States and Perturbative Description at $\Delta_1$}  \label{sec_Delta1}

Let us first focus on the divisor $\Delta_1$ of type II$_1$. We begin by discussing the light states in the vicinity of $\Delta_1$ in the original Type IIB formulation and subsequently analyse the heterotic dual formulation. 

\subsection{Type IIB Perspective}
According to the general discussion in~\cite{Friedrich:2025gvs},  along $\Delta_1$ the K3 surface $Z$ appearing in \eqref{Ydegeneration} has a transcendental lattice of signature ${\rm sgn}(\Lambda_{\rm trans})=(2,1)$. Two kinds of states become light in the vicinity of $\Delta_1$: BPS states with charges $q\in H^3(Y,\mathbb{Z})\cap {\rm Gr}_2(\Delta_1)$ and a string obtained as a supergravity solution that realises the limit $x_1\to 0$ at the string core, dubbed EFT string in~\cite{ Lanza:2020qmt,Lanza:2021udy}. As discussed in~\cite{Friedrich:2025gvs}, the tension of the string scales in the limit $x_1\to 0$ as  
\begin{equation} \label{tensionT}
    \frac{T}{M_{\rm Pl}^2} \sim e^{K} = \frac{1}{-\frac{1}{4\pi^3}\log |x_1| +\dots }\,,
\end{equation}
where $K$ is the K\"ahler potential that can be computed from the periods in~\eqref{eq:periods} as 
\begin{equation} \label{expK}
   e^{-K}= -i \left(\bar{X}^I F_I - X^I \bar{F}_I\right) = -\frac{1}{4\pi^3} \log |x_0^2x_1|+\mathcal{O}(x_0)\,. 
\end{equation}
The BPS states with charge $q\in H^3(Y,\mathbb{Z}) \cap {\rm Gr}_2(\Delta_1)$ are obtained from D3-branes wrapping special Lagrangian three-cycles that shrink in the limit $x_1\to 0$. By making explicit use of the degenerate geometry~\eqref{Ydegeneration}, these sLag three-cycles $\Gamma$ can be identified as $S^1$-fibrations over curves $C\in \Lambda_{\rm trans}(Z)$~\cite{Friedrich:2025gvs}. Here, the fibre is the $S^1$ transverse to $Z$ that shrinks at the degenerate locus $y=0$. The calibration of $\Gamma\in H_3(Y)$ is then inherited from the calibration of $C\in \Lambda_{\rm trans}(Z)$. As $\text{sgn}(\Lambda_{\rm trans}) =(2,1)$, there is a basis of curves $C_i\in \Lambda_{\rm trans}$, $i=0,1,2$, for which 
\begin{equation}\label{eta1}
    \eta_{ij} \equiv C_i\cdot_Z C_j =\begin{pmatrix}
        2&0&0\\0&0&1 \\ 0&1&-2
    \end{pmatrix}\,, \quad  i,j \in \{0,1,2 
  \}  \,.
\end{equation}
The lattice of three-cycles shrinking at $x_1=0$ is thus three-dimensional and spanned by $S^1 \hookrightarrow \Gamma_i\to C_i$. The periods 
\begin{equation}
    \Pi^i=\int_{\Gamma_i} \Omega_3 \,,
\end{equation}
of these three-cycles are a linear combination of $X^{0,1,2}$ appearing in~\eqref{eq:periods} and determine the BPS masses of the D3-branes wrapped along $\Gamma_i$ via
\begin{equation} \label{BPSmass}
\frac{M_{\rm BPS}}{M_{\rm Pl}} = e^{K/2} \left|\int_{\Gamma_i} \Omega_3 \right|\,.
 \end{equation}
 
For generic values of $x_0$ along $\Delta_1$, these masses vanish uniformly for all $\Gamma_i$, $i=0,1,2$, because of the overall suppression by $e^{K/2}$ following from (\ref{expK}) \cite{Grimm:2018cpv,Grimm:2018ohb}, independently of the value of the periods. This reflects the vanishing of the radius of the $S^1$ fiber of $\Gamma_i$ \cite{Friedrich:2025gvs}.

Along $\Delta_1$ the parameter $x_0$ can be chosen freely and parametrizes the ratios of the periods of the transcendental curves $C_i$ on $Z$, i.e., it parameterizes the complex structure moduli space of the polarized K3-surface $Z$. As is clear from the intersection matrix $\eta_{ij}$ and the signature of $\Lambda_{\rm trans}$, the curve $C_2$ has self-intersection $(-2)$ and is hence shrinkable inside $Z$. Accordingly, there exists a locus in the complex structure moduli space of $Z$ at which $C_{2}$ shrinks to a point. Since the three-cycle $\Gamma_2$ is an $S^1$-fibration over $C_{2}$, its period also vanishes if $C_2$ shrinks to a point. Comparing with the periods in~\eqref{eq:periods} we can identify the locus $x_0=0$ as the conifold locus for the K3-surface $Z$ at which $C_2$ shrinks to a point. Accordingly, the period of $\Omega_3$ over $\Gamma_2$ is given by
\begin{equation}\label{Periods-2}
    \int_{\Gamma_2} \Omega_3 =X^2 = -\frac{\beta}{\pi}\sqrt{x_0}\,.
\end{equation}
Instead, for $x_0=0$, the curves $C_{0,1}$ remain of finite size inside $Z$. This implies that the periods $\Pi^{0,1}$ receive a non-trivial contribution from $X^0$ and/or $X^1$ (possibly along with $X^2$). 

To summarize, at generic points along $\Delta_1$, the light spectrum of states is determined by a tensionless BPS string and BPS particles of mass 
\begin{equation}
    \frac{M_{\rm BPS}}{M_{\rm Pl}} \sim \frac{1}{\sqrt{-\frac{1}{4\pi^3}\log |x_1|}} \,. 
\end{equation}
In fact, as argued in~\cite{Friedrich:2025gvs}, there is an infinite tower of such BPS states from the two-dimensional sublattice of 3-cycles fibered over curves in $\Lambda_{\rm trans}$ of non-negative self-intersection on $Z$. Furthermore, as we approach the intersection $\Delta_1\cap \Delta_0$ by sending $x_0\to 0$, in addition to this tower, a \emph{single} BPS state becomes light as  
\begin{equation}
    \frac{M_{-2}}{M_{\rm Pl}}= \frac{\frac{\beta}{\pi} \Big|\sqrt{x_0}\Big|}{\sqrt{-\frac{1}{4\pi^3}\log|x_1|}}+\dots\,, 
    \end{equation}
obtained from a D3-brane wrapping the $S^1$-fibered three-cycle $\Gamma_2$ whose base two-cycle $C_2$ shrinks. The vanishing locus for $C_2$ corresponds to a gauge enhancement point at which a $U(1)$ gauge theory enhances to $SU(2)$. 
 Along $\Delta_1$, one can identify three $U(1)$
 gauge factors, corresponding to the elements in ${\rm Gr}_2(\Delta_1)$, which are weakly coupled in the sense  that the ratio $\Lambda_{\rm WGC}/\Lambda_{\rm sp} \lesssim 1$ \cite{Cota:2022maf}. Here $\Lambda_{\rm WGC} = g_{\rm YM} M_{\rm Pl}$ and $\Lambda_{\rm sp}$ is the species scale \cite{Dvali:2007hz,Dvali:2009ks}, set by the tension of the BPS string, (\ref{tensionT}).    What happens in the regime $x_1\ll x_0\ll 1$ is that one of these $U(1)$s, namely the one which enhances to $SU(2)$, effectively decouples from the gravitational sector. This can be seen by noticing that the central charge of the D3-brane on $\Gamma_2$  vanishes, which implies that the state is not charged under the gravi-photon. In other words, the gravi-photon does not receive a contribution from the $U(1)$ photon that couples to the D3-brane on $\Gamma_2$. In the vicinity of $\Delta_0\cap \Delta_1$, the prepotential of the full $\cN=2$ supergravity theory is given by 
\begin{equation}\begin{aligned}
    F = F_I X^I &=\frac{\beta^2}{\pi^3 i} x_0 \log(x_0) \\
    &-\frac{\beta^2x_0}{72\pi^3 i} \left(5 x_0 -\frac{9x_1}{2}\right) 
    \log (x_1) 
 +\dots \,.  \label{prepot-1a}
\end{aligned}\end{equation}
 This follows from the periods as expanded in Appendix \ref{app-periods}.
 In the regime $x_1\ll x_0$, (\ref{prepot-1a}) can be approximated as 

\begin{equation}\label{Fapprox1}
    F \stackrel{x_1\ll x_0}{\simeq }-\frac{5\beta^2}{72\pi^3 i} x_0^2 \log x_1 \,.
\end{equation}

 Recalling that ${\rm log}(x_0)$ and ${\rm log}(x_1)$ are flat local coordinates, we notice that the logarithmic term in $x_1$ signals that $x_1$ controls a gravitational sector that becomes weakly coupled as $x_1\to 0$. Instead, $F$ being polynomial in $x_0$ indicates that $x_0$ controls a field theory sector that is decoupling from gravity in this regime in the sense that the associated $U(1)$ gauge factor does not contribute to the gravi-photon, as discussed above.

\subsection{Dual Heterotic Perspective} 
So far, our discussion proceeded in terms of the explicit geometry of $Y$ as probed by compactifying Type IIB string theory thereon.
 As expected from the Emergent String Conjecture~\cite{Lee:2019oct}, in the vicinity of $\Delta_1$ the theory allows a weakly coupled dual description. As demonstrated in~\cite{Friedrich:2025gvs}, this dual description is associated with the BPS string that becomes tensionless at $\Delta_1$ as in (\ref{tensionT}). By explicitly counting the worldsheet degrees of freedom on this string, \cite{Friedrich:2025gvs} established that this string is a critical heterotic string. More precisely, for the type II$_1$ degenerations,  the  tensionless string can be identified with the critical heterotic string compactified on K3$\times T^2$, where the radius of one of the one-cycles, $\gamma_1\in H_1(T^2)$, is fixed to its self-dual value. Let us stress that the heterotic K3 and the K3 surface $Z$ appearing in the degeneration~\eqref{Ydegeneration} are different. A gauge background for the enhanced gauge group at the self-dual radius breaks the original gauge group $U(1)^4$ associated with the two Kaluza--Klein and the two winding $U(1)$s on $T^2$ to $U(1)^3$. Similarly, the $E_8 \times E_8$ part of the heterotic gauge group is broken by the gauge background. This reproduces the weakly coupled $U(1)^3$ gauge group encountered also in the Type IIB description along a generic point on $\Delta_1$.

The above interpretation is consistent with the mirror dual picture developed in~\cite{Kachru:1995fv}. The complexified heterotic string coupling in this regime can be identified as 
\begin{equation}\label{id:Sx1}
    S = \frac{1}{2\pi i} \log x_1 +\dots \,. 
\end{equation}
The radius of the other one-cycle, $\gamma_2\in H_1(T^2)$, is instead controlled by $x_0$, with $x_0\gg 1$ corresponding to the large radius regime. The BPS states obtained by wrapping D3-branes on $\Gamma_{1,2}$ map to winding and momentum modes of the fundamental heterotic string along $\gamma_2$. By contrast, the D3-brane on $\Gamma_0$ maps to a linear combination of winding and momentum states of the fundamental heterotic string along $\gamma_1$. It is charged under the linear combination of KK and winding $U(1)$s associated with $\gamma_1$ which is left unbroken by the gauge background.

Given the above identification, the interpretation of the regime $x_0\ll1$ in the heterotic dual description is straightforward: The D3-brane state wrapping $\Gamma_2$ that becomes massless at $x_0=0$ corresponds to the heterotic string state with winding and momentum number $(1,1)$ along $\gamma_2$. Recall that the mass of a heterotic string excitation at left-moving excitation level $N_L=0$ with winding and momentum mode $(w,k)$ and vanishing charge under the $E_8\times E_8$ gauge group is given by 
\begin{equation}\label{masshetex}
    \alpha'_{\rm het}M_{(w,k)}^2 = \frac{\alpha'_{\rm het}}{2} \left(\frac{k }{R_{\gamma_2}} + \frac{wR_{\gamma_2}}{\alpha'_{\rm het}} \right)^2 - 2 \,,
\end{equation}
where the $-2$ comes from the left-moving vacuum energy of the heterotic string worldsheet and $R_{\gamma_2}$ is the radius of $\gamma_2$.

As is well-known, the state with $(w,k)=(1,1)$ becomes massless at the self-dual radius for $\gamma_2$. The point $x_0=0$ thus maps in the heterotic theory to 
$R_{\gamma_2} M_{\rm het}=1$
with $M_{\rm het} = 1/\sqrt{\alpha_{\rm het}'}$.
For $(w,k)=(1,1)$, we can then equate \eqref{masshetex} with the square of \eqref{Periods-2}, such that for $x_1\ll x_0\ll 1$ we arrive at the approximate identification
\begin{equation}\label{id:x0R}
    \frac{\alpha'_{\rm het}}{2} \left(\frac{1 }{R_{\gamma_2}} + \frac{ R_{\gamma_2}}{\alpha'_{\rm het}} \right)^2 - 2 = \frac{\beta ^2}{\pi^2} |x_0| \,.
\end{equation}
This state can be interpreted as the W-boson enhancing a $U(1)$ gauge theory, given by a linear combination of heterotic winding and KK $U(1)$s, to $SU(2)$.

Notice that along the divisor $\Delta_1$, the states with central charges $F_0, F_1,F_2$ 
become infinitely heavy. They are obtained from D3-branes wrapping the cycles $\Gamma^{i}$ dual to $\Gamma_{i}$ for $i=0,1,2$. In the heterotic dual theory, these can be interpreted as states obtained from the heterotic NS5-brane. In particular, for later reference, we note that along $\Delta_1$ the D3-brane wrapping $\Gamma^2$ with central charge $F_2$ corresponds to an NS5-brane wrapping K3$\times \gamma_2$ in the heterotic dual theory. 

The prepotential~\eqref{Fapprox1} must be contrasted to the prepotential in the regime $x_1\ll x_0^{-1}\ll 1 $,
\begin{equation}
    F \sim \log x_1 (\log x_0)^2\,,
\end{equation}
which is polynomial in the flat coordinate $t_i=\log x_i$ as opposed to the exponential dependence in \eqref{Fapprox1} on $t_0 = \log x_0$. The difference between the polynomial and exponential prepotential becomes clear in the heterotic dual: If the heterotic dual has a large-radius supergravity description, the prepotential is polynomial (as expected from supergravity). If instead the heterotic dual is compactified on a string-size manifold, the leading order prepotential contains exponential terms in $t_0$. Notice that along $\Delta_1$ no exponential terms in the heterotic string coupling $S$ given by (\ref{id:Sx1}) appear at leading order. This reflects the fact that the heterotic dual is fully perturbative in the string coupling and in particular does not contain any spacetime-filling NS5-branes.

\section{Light States and Perturbative Description at $\Delta_0$}  \label{sec_Delta0}

We now turn to the divisor $\Delta_0$ associated, at generic points, with a type II$_0$ singularity. In the following, we thus focus on the regime $x_0\ll x_1 \ll 1$. As we will see, the physical interpretation differs in some aspects from the regime $x_1\ll x_0 \ll 1$ discussed above. We begin with the original Type IIB perspective to identify the light states in the vicinity of $\Delta_0$ and then move to the heterotic dual. 
\subsection{Type IIB Perspective} 
Since along the locus $\Delta_0$ in moduli space, 
 the Calabi-Yau $Y$ develops  a type II$_0$ singularity of Tyurin type, we can again apply the results of~\cite{Friedrich:2025gvs} to identify light states in the vicinity of $\Delta_0$. As before, there is a string whose tension scales in the vicinity of $x_0=0$  as
\begin{equation}\label{tension0}
    \frac{T}{M_{\rm Pl}^2} \sim e^K = \frac{1}{-\frac{1}{2\pi^3}\log |x_0|+\dots}\,. 
\end{equation}
The spectrum of light BPS states at a generic point along $\Delta_0$ differs, however, from the spectrum along generic points on $\Delta_1$. The reason is that at generic points on $\Delta_0$, the space $\text{Gr}_2(\Delta_0)$ is only two-dimensional, with the missing element lying now in ${\rm Gr}_3(\Delta_0)$. This is a result of the geometric transition interpolating between the loci $\Delta_1$ and $\Delta_0$ in moduli space. Geometrically, this means that the transcendental lattice of the K3 surface $Z$ in~\eqref{Ydegeneration} has signature ${\rm sgn}(\Lambda_{\rm trans}(Z))=(2,0)$. Accordingly, there is only a two-dimensional lattice of $S^1$-fibred special Lagrangian 3-cycles in the vicinity of $\Delta_0$, $S^1\hookrightarrow \widetilde\Gamma_{0,1} \to \widetilde C_{0,1}$. Here, $\widetilde C_{0,1}$ are the generators of $\Lambda_{\rm trans}(Z)$ given by 
\begin{equation}
    \widetilde{C}_0=C_0\,,\quad \widetilde{C}_1=2C_1+C_2\,,
\end{equation}
in terms of the curves $C_i$ introduced above~\eqref{eta1}. The curves $\widetilde{C}_{0,1}$ satisfy
\begin{equation}
    \widetilde \eta_{ij}  \equiv \widetilde{C}_i \cdot_Z \widetilde{C}_j = \begin{pmatrix}
        2&0\\0&2 
    \end{pmatrix}\,. 
\end{equation}
The central charges of D3-branes wrapping $\widetilde{\Gamma}_0$ and $\widetilde{\Gamma}_1$ are constant in the limit $x_0\to 0$ and hence are linear combinations of the periods in~\eqref{eq:periods} containing $X^0$ and/or $X^1$. Importantly, at generic points along $\Delta_0$, the K3 surface $Z$ is non-singular and hence does not contain any calibrated cycles that are shrunk to zero size. According to the analysis in~\cite{Friedrich:2025gvs}, all BPS states with charge $q\in {\rm Gr}_2(\Delta_0)$ correspond to D3-branes wrapping special Lagrangian cycles that are $S^1$-fibrations over calibrated curves in $Z$. Since along $\Delta_0$ the K3 surface $Z$ does not contain any calibrated curves of vanishing size, there are also no BPS states whose central charge vanishes identically at a generic point along $\Delta_0$. As discussed in~\cite{Hattab:2025aok} in the context of K-points, the mass of such a state, if it existed as a BPS state, would decay faster than the mass of a generic state with charge in ${\rm Gr}_2(\Delta_0)$ as we approach $\Delta_0$, i.e., super-exponentially in the moduli space distance. 

As noted already, unlike for $\Delta_1$, the space ${\rm Gr}_3(\Delta_0)$ is non-empty. The states with charge $q\in H^3(Y,\mathbb{Z})\cap  {\rm Gr}_3(\Delta_0)$ also become massless along $\Delta_0$ and correspond to a field theory sector that decouples from gravity. In general, the relation between the secondary type and the rank of the decoupled gauge group has been discussed in general in~\cite{Marchesano:2024tod,Castellano:2024gwi,Monnee:2025ynn} and the details of the decoupling of a strongly coupled field theory sector at singularities of the type discussed here have been analyzed in~\cite{Castellano:2024gwi}. 

To understand in what sense there is a decoupling gauge sector, note that in our example, ${\rm Gr}_3(\Delta_0)$ is two-dimensional with the corresponding $U(1)$ gauge groups being electric-magnetic duals of each other. The states charged under these $U(1)$s arise from D3-branes wrapping sLag 3-cycles $\Gamma_2$ and $\Gamma^2$, and we identify their periods as
\begin{equation} \label{WMonoperiod}
    \left(a, a_D\right) \equiv \left(\int_{\Gamma_2}\Omega_3,\int_{\Gamma^2}\Omega_3 \right)  = \left(X^2, F_2\right)\,.  
\end{equation}
As discussed in~\cite{Kachru:1995fv}, these states can be identified with the W-boson and magnetic monopole of an $\cN=2$ $SU(2)$ super-Yang--Mills theory. Thus, the $U(1)$s associated with elements in ${\rm Gr}_3(\Delta_0)$ form a decoupled Seiberg--Witten field theory sector.  The local coordinate $x_1$ now controls the ratio between electric and magnetic periods as 
\begin{equation} \label{aDa}
    \frac{a_D}{a} =- \frac{1}{2\pi i} \log x_1 - \frac{2}{i\pi} +\dots\,. 
\end{equation}
For $x_1\to 0$, the magnetic monopole becomes infinitely heavy relative to the W-boson and we obtain a purely electric gauge theory. The local coordinate $x_0$ controls the bare coupling of the SYM theory. To see this, we notice that the prepotential 
\begin{equation}\label{Fapprox0}
    F = \frac{\beta^2}{\pi^3 i}x_0 \log x_0 + \dots\,,
\end{equation}
can be identified with the Seiberg--Witten prepotential~\cite{Seiberg:1994rs} for which the heterotic string coupling $\widetilde{S}$, see equation \eqref{eq:S-tilde}, determines the bare coupling.

As is manifest from (\ref{Fapprox0}), the prepotential contains logarithmic and polynomial expressions in $x_0$. The logarithmic expression signals the emergence of a weakly coupled gravitational description, whereas the polynomial term reflects that a field theory sector decouples from gravity. Again, the $U(1)$ gauge theory decouples from gravity since it does not contribute to the gravi-photon. In addition, and unlike along $\Delta_1$, at generic points along $\Delta_0$ the $U(1)$ gauge theory is also strongly coupled. This follows from the fact that $\Lambda_{\rm WGC}/\Lambda_{\rm sp} \gg 1$ \cite{Cota:2022maf}  because the gauge coupling is controlled by the Hodge norm of an element of $q \in {\rm Gr}_3(\Delta_0)$, rather than in ${\rm Gr}_2$, and for these $||q||^2 \sim {\cal O}(1)$.
 Notice that in contrast to~\eqref{Fapprox1}, here both logarithmic and polynomial terms depend on $x_0$. This signals that from the perspective of the emerging weakly coupled gravitational theory, the field theory subsector is non-perturbative. We will propose a heterotic dual interpretation of this in the following.

\subsection{Dual Heterotic Perspective}
 The string becoming tensionless along $\Delta_0$ as in~\eqref{tension0} is a critical heterotic string \cite{Friedrich:2025gvs}. In the regime $x_0\ll x_1$, we can thus identify the heterotic string coupling as
\begin{equation} 
\label{eq:S-tilde}
\widetilde{S} = \frac{1}{2\pi i} \log x_0 +\dots\,. 
\end{equation} 
Furthermore, the dual heterotic string is compactified on K3$\times T^2$ with the radius of both one-cycles $\gamma_{1,2}\in H_1(T^2)$ fixed to the self-dual value \cite{Friedrich:2025gvs}. This reflects the rigidity of the K3 surface $Z$ in the Type IIB picture along $\Delta_0$. Compared to the heterotic string emerging at generic points along $\Delta_1$, one thus expects a gauge enhancement $U(1)\to SU(2)$ for a linear combination of the winding and KK $U(1)$ associated with $\gamma_2$. However, along $\Delta_0$, the resulting $SU(2)$ gauge theory is completely Higgsed by a non-Abelian gauge bundle. This can be seen directly from the worldsheet theory of the EFT string associated with $\Delta_0$ \cite{Friedrich:2025gvs}. The spacetime $U(1)$ gauge theory in question is a perturbative heterotic gauge theory and thus corresponds to a left-moving worldsheet algebra associated with a half-Fermi multiplet of the 2d $\cN=(0,4)$ worldsheet theory. The analysis of~\cite{Friedrich:2025gvs} shows that this multiplet has a non-trivial interaction on the worldsheet which is associated with a gauge bundle in spacetime. In fact, the only free fields on the worldsheet giving rise to spacetime $U(1)$s are right-moving scalars. Thus the perturbative heterotic gauge group for the string emerging at $\Delta_0$ is $U(1)^2$. 

In particular, this means that the electric and magnetic $U(1)$s associated with elements in ${\rm Gr}_3(\Delta_0)$ do not have a perturbative heterotic origin in the regime $x_0\ll x_1$. This is no surprise since already along $\Delta_1$ the state with central charge given by $F_2$ arises from a heterotic NS5-brane wrapping K3$\times \gamma_2$. In the regime $x_0\ll x_1$, we saw above that the states with central charge $(a,a_D)$ form the electric and magnetic states of $SU(2)$ Seiberg--Witten theory. The associated electric and magnetic $U(1)$s must be of non-perturbative origin in the heterotic theory. For compactifications of the heterotic string to four dimensions, non-perturbative $U(1)$ gauge theories can arise from spacetime-filling NS5-branes wrapping elliptic curves in the compact manifold. Since these $U(1)$s arise from the tensor on the NS5-brane worldvolume theory, they are automatically decoupled from gravity in the weak string coupling regime. Our proposal is therefore that from the heterotic perspective, the $U(1)$ gauge theories associated with elements in ${\rm Gr}_3(\Delta_0)$ in fact arise from spacetime-filling NS5-branes that, as we will see momentarily, wrap the heterotic torus.

Along $x_0=0$, the parameter $x_1$ maps to the position of the NS5-brane along the Ho\v{r}ava--Witten interval in the heterotic duality frame. The states with charge $q\in H^3(Y,\mathbb{Z})\cap {\rm Gr}_3(\Delta_0)$ now correspond to two E-strings wrapped on $\gamma_2$. For the E-string to wrap a one-cycle in the heterotic $T^2$, the NS5-brane containing this E-string must wrap the $T^2$ as well, as proposed above. In the Ho\v{r}ava--Witten picture, the two E-strings $\mathtt{E}_{1,2}$ correspond to M2-branes ending on the NS5-brane and either of the 9-branes at the ends of the interval~\cite{Ganor:1996mu,Seiberg:1996vs}. The tension of these strings is given by 
\begin{equation}
    T_{\mathtt{E}_1} = \kappa M_{\rm het}^2 \,,\quad T_{\mathtt{E}_2}=(1-\kappa) M_{\rm het}^2\,,
\end{equation}
where $\kappa$ parameterizes the location of the NS5-brane along the HW interval. The mass of a wrapped E-string with wrapping number $w$ is
\begin{equation}\label{eq:MEclassical}
    M^2_{\mathtt{E}_{1,2}, w} = T_{\mathtt{E}_{1,2}}\left(\frac{w^2}{2} R_{\gamma_2}^2 M_{\rm het}^2 -\frac12 \right)\,,
\end{equation}
where we used that the E-string has vacuum energy $E_0~=~1/2$~\cite{DelZotto:2016pvm,Kim:2018gak}. We see that the once-wrapped E-string ($w=1$) hence gives a massless state for the self-dual radius $R_{\gamma_2}^2 M_{\rm het}^2 =1$. For E-strings arising in 6d F-theory compactifications, this can be also be seen explicitly using F-/M-theory duality~\cite{Klemm:1996hh}. In four dimensions, the expression~\eqref{eq:MEclassical} can be corrected by non-perturbative effects, i.e. 
\begin{equation}
       M^2_{\mathtt{E}_{1,2}, w} = T_{\mathtt{E}_{1,2}}\left(\frac{w^2}{2} R_{\gamma_2}^2 M_{\rm het}^2 -\frac12 + \delta_{1,2}\right)\,,
\end{equation}
where $\delta_{1,2}\sim \mathcal{O}\left(e^{2\pi i \widetilde{S}}\right)$. These corrections are computable using the dual Type IIB picture since the winding states of the E-strings are identified with the $W$-boson and monopole of the $SU(2)$ SW theory. The periods entering the computation of their masses in Planck units via (\ref{BPSmass}) are given in (\ref{WMonoperiod}). We summarize the interpretation of the states in the different heterotic duality frames along $\Delta_0$ and $\Delta_1$ in Table~\ref{tab:summary}. Using this dictionary, along with the fact that the ratio between $M_{\rm het}$  and $M_{\rm Pl}$ is set by (\ref{tension0}), we match 
\begin{eqnarray}
    \frac{T_{\mathtt{E}_1}}{M_{\rm het}^2} \delta_1 &=& \frac{\beta^2}{\pi} |x_0| +\dots  \,,\\  \frac{T_{\mathtt{E}_2}}{M_{\rm het}^2} \delta_2 &=& \frac{4\beta^2}{\pi^2}  |x_0|  \left|\frac14\log(x_1)+ 1 \right|^2 +\dots \,,
\end{eqnarray}
such that for $\kappa\ll 1$ we can identify 
\begin{equation*}
    \delta_{1,2} = \frac{\beta^2}{\kappa}e^{-2\pi \,\text{Im}(\widetilde{S}) }\,,\quad \frac{1-\kappa}{\kappa} = \frac{4}{\pi^2}  \left|\frac{1}{4}\log(x_1)+ 1  \right|^2\,. 
\end{equation*}
In other words, in the regime $x_0\ll x_1$, the coordinate $x_1$ parametrizes the location of the spacetime-filling NS5-brane along the HW interval. In heterotic variables, the prepotential~\eqref{Fapprox0} reads 
\begin{equation} \label{Fhet}
    F=  \frac{2\beta^2}{\pi^2}\widetilde S \exp(2\pi i \widetilde{S}) +\dots\,. 
\end{equation}
This form of the prepotential reflects the fact that the gravitational duality regime corresponds to a weakly coupled heterotic string theory, but that there is a decoupled field theory sector that is non-perturbative from the heterotic perspective. Crucially, this sector arises from spacetime-filling NS5-branes and is non-perturbative in the heterotic string coupling, as indicated by the exponential term in $\widetilde{S}$ appearing at leading order in the prepotential. This is to be contrasted to the exponential terms appearing in the leading terms in the prepotential (\ref{Fapprox1}) for $x_1\ll x_0\ll 1$, which give non-perturbative corrections in $\alpha'_{\rm het}$. 

Let us stress that the heterotic string is weakly coupled at any energy scale below $M_{\rm het}$ and that it gives rise to the full \emph{gravitational} theory. In particular all states predicted by the Distance Conjecture~\cite{Ooguri:2006in} arise as excitations of the heterotic string as expected from the Emergent String Conjecture~\cite{Lee:2019oct}. The non-perturbative part is simply a field theory  sector that decouples from the gravitational theory. The presence of the decoupled sector associated with a spacetime filling NS5-brane is an indication that there is no straightforward CFT description capturing this sector in the heterotic frame. Moreover, weak-coupling limits for the heterotic string with NS5-brane sectors are in no way exotic. In six dimensions they are well-studied and correspond to heterotic strings with Little String Theory sectors. 
 Emergent string limits including such sectors are realised in F-theory compactifications whose base are a blow-up of a Hirzebruch surface $\mathbb F_n$, and have been discussed in this context in detail already in~\cite{Lee:2018urn}. The presence of the NS5-brane is reflected in a holomorphic anomaly of the elliptic genus of the otherwise weakly coupled heterotic string. 
 Thus, from the perspective of the Distance and Emergent String Conjectures, type II$_0$ limits are in no way more special than these constructions. In particular, we are finding no indication that a refinement of the Distance Conjecture is required to understand these kinds of limits. As pointed out in \cite{Hattab:2025aok}, such a refinement would be implied if there were  additional super-exponentially light states with charge $q\in H^3(Y,\mathbb{Z})\cap {\rm Gr}_2$ at generic points of $\Delta_0$.
 
 \begin{table*}

    \renewcommand{\arraystretch}{1.3}
    \begin{tabular}{|c|c|c|c|}
\hline
           \textbf{D3-Brane charge} &\textbf{D3-Brane charge} &\textbf{Heterotic Dual} & \textbf{Heterotic Dual}  \\
        $q\in {\rm \textbf{Gr}}_\ell(\Delta_1)$&$q\in {\rm \textbf{Gr}}_\ell(\Delta_0)$&\textbf{along} $\Delta_1$ & \textbf{along} $\Delta_0$\\ \toprule
         \multicolumn{2}{|c|}{$\ell=2$}&\multicolumn{2}{c|}{generic winding and momentum of het. string on $\gamma_1$ and $\gamma_2$} \\\hline
        $\ell=2$&$\ell=3$ &(winding,momentum)$=(1,1)$ of het. string on $\gamma_2$ &winding of $\mathtt{E}_1$ on $\gamma_2$\\ \hline 
        $\ell=4$ & $\ell=3$& NS5-brane on K3$\times \gamma_2$& winding of $\mathtt{E}_2$ on $\gamma_2$\\ \hline
    \end{tabular}
    \caption{Summary of light BPS States and their heterotic interpretation along $\Delta_0$ and $\Delta_1$. The second and third line describe the W-boson and, respectively, the monopole of a Seiberg--Witten subsector.\label{tab:summary}}
\end{table*}

Let us close the investigation of the two-modulus example with a comment. In the discussion above, we have treated the two regimes $x_1\ll x_0$ and $x_0\ll x_1$ separately, finding that their heterotic duals differ by the presence of spacetime-filling NS5-branes and a non-Abelian gauge bundle for one of the perturbative heterotic gauge groups. Since both regimes are part of the same moduli space,  there should exist a smooth transition between the two perturbative heterotic descriptions occurring for $x_1\sim x_0 \ll 1$. This in particular includes the locus $\Delta_0\cap \Delta_1$. From the heterotic perspective, it seems that in this transition, we nucleate a pair consisting of a spacetime-filling NS5-brane and the non-Abelian gauge bundle for the heterotic string on K3$\times T^2$. 
 Such a pair nucleation is reminiscent of the process proposed in \cite{Anderson:2022bpo}, though this analysis is for conifold transitions of Calabi--Yau threefolds in heterotic 4d $\cN=1$ compactifications.
 While the details of the nucleation process remain to be understood better, already the Type IIB dual provides an exact description of the full non-perturbative (in $g_s$) heterotic string theory,  implying that such a pair production process should be possible in the full non-perturbative heterotic string theory in 4d. 
 In Type IIB, the way how the non-perturbative $U(1)$ along $\Delta_0$ (described by an element in Gr$_3(\Delta_0)$) becomes perturbative along the enhancement from II$_0$ to II$_1$ (where it corresponds to an element in Gr$_3(\Delta_1)$) was explained generally in \cite{Monnee:2025ynn}.
 Understanding this process from the heterotic perspective would be very interesting. We stress, however, that our discussion above of the physics in the vicinity of the type II$_0$ divisor does not rely on the microscopics of this transition. 

\section{From $\Delta_0$ to the K-point} \label{sec_DeltatoK}
Starting from the type II$_0$ divisor in the two-modulus example, we 
propose to consider an extremal transition to a one-parameter model with a K-point. From a field theory point of view, such a transition should be possible by Higgsing the massless monopole/anti-monopole pair.\footnote{It would be interesting to construct the model after the transition explicitly. Note that it may not admit a simple geometric description, in particular in view of the fact that on the mirror dual Type IIA side, the massless state corresponds to a wrapped D6-brane. In \cite{Doran:2024kcb}  conifold transitions from 2-parameter models were analysed, including a transition in the present geometry by Higgsing massless modes arising at the intersection of the II$_1$ divisor with an I$_1$ divisor. The Higgsed model, however, does not admit a K-point after the transition. By contrast, the transition we have in mind would start from the intersection of the II$_0$ divisor with a different I$_1$ branch of the discriminant.}

K-points are type II$_0$ singularities in one-parameter Calabi--Yau threefolds \cite{vanstraten2017calabiyauoperators,Joshi:2019nzi}. In the recent work~\cite{Hattab:2025aok}, these points have been investigated in detail in the context of the Emergence Proposal \cite{Grimm:2018ohb,Palti:2019pca}. In particular, as stressed in~\cite{Hattab:2025aok}, in the vicinity of a K-point the prepotential takes the form
\begin{equation}\label{FKK}
    F_K \sim 
    x_0 \log x_0 + \dots\,,
\end{equation}
where $x_0$ is the local coordinate around the K-point located at $x_0=0$. In~\cite{Hattab:2025aok}, this form of the prepotential was interpreted as potential evidence that the K-point cannot be described by a heterotic string that is weakly coupled all the way to the string scale. 
 Indeed, our analysis from the previous sections suggests non-perturbative physics near K-points, but in the form of an NS5-brane sector in an otherwise weakly coupled perturbative heterotic string governing the perturbative gravitational duality frame up to the heterotic string scale.

To see this, we first notice that, as expected, the prepotential in~\eqref{FKK} agrees with the leading order prepotential along the divisor $\Delta_0$ in the two-parameter model analyzed previously. In that model, we concluded that there is a weakly coupled description in terms of a heterotic string that is weakly coupled all the way to the string scale, with the polynomial expression in $x_0$ merely indicating the presence of a non-perturbative field theory sector that is decoupled from the gravitational theory. In order to argue that a similar conclusion also holds for K-points in one-modulus examples, we now connect the two-parameter model discussed above to a one-parameter model via an extremal transition. 

Consider again the singular divisor $\Delta_0$ corresponding to $x_0=0$. As illustrated in Figure \ref{fig:duality-frames}, at $x_1\sim {\rm exp}(-4)$, this divisor intersects a conifold locus along which the masses of the monopole and the W-boson of the $SU(2)$ SW-theory decouple as
\begin{equation}
    \frac{a_D}{a} = 0\,,
\end{equation}
see (\ref{aDa}). We can thus write down a local field theory for the magnetic $U(1)$ coupled to a massless monopole which is part of a hypermultiplet. By contrast, the W-boson of the electric $SU(2)$ that becomes massless with respect to the string scale at the point $x_1=0$ is part of a vector multiplet. Thus, by giving a VEV to the monopole, we can Higgs the magnetic $U(1)$, such that we are left with a theory with a single massless vector multiplet corresponding to the remaining weakly coupled $U(1)$ (in addition to the weakly coupled gravi-photon). This theory corresponds to a one-parameter model with a K-point at $x_0$ and prepotential as in~\eqref{FKK}. The transition that we described by giving a VEV to the monopole does not affect the gravitational theory in any way, which is still best described by a weakly coupled heterotic string. In particular, the BPS states with charges in $q\in H^3(Y,\mathbb{Z}) \cap {\rm Gr}_2$ are the same as before. This implies that there is no BPS state with mass proportional to $x_0$. Still, the theory contains a field theory subsector (corresponding to a massless hypermultiplet coupled to a Higgsed gauge field) that does not have a perturbative heterotic origin as it arises from a spacetime-filling NS5-brane. The presence of the spacetime-filling NS5-brane manifests itself in the leading term in the prepotential containing exponential expressions in the heterotic dilaton $\widetilde{S}=1/(2\pi i) \log x_0$. We reiterate that the emergence of a weakly coupled heterotic string compactified on a background containing spacetime-filling NS5-branes is in perfect agreement with the Emergent String Conjecture. 

\section{Discussion} \label{sec_disc}
In this note we have scrutinized the physics of 
type II$_0$ degenerations and, as a special case, K-points in the complex structure moduli space of Type IIB string theory on Calabi--Yau threefolds.
 Our main conclusions support the picture developed in \cite{Friedrich:2025gvs, Monnee:2025ynn}, according to which
 the physics of type II$_0$ degenerations is in perfect agreement with the conventional interpretation of both the Distance Conjecture and the Emergent String Conjecture. 
 Extrapolating our findings from the concrete two-parameter model of \cite{Candelas:1993dm,Kachru:1995fv} studied in this note, the physics associated with a type II$_0$ degeneration in the moduli space of a Calabi--Yau threefold with $h^{2,1}> 1$ can be summarised as follows: Near the type II$_0$ degeneration, the asymptotic theory contains a weakly coupled heterotic string, two weakly coupled $U(1)$ sectors (encoded in the two elements of  Gr$_2$ characteristic of a II$_0$ degeneration \cite{Grimm:2018cpv,Grimm:2018ohb}) and one or more non-perturbative field theory sectors. The latter decouple from gravity and are associated, together with their magnetic duals, with the elements of Gr$_3$. In the dual heterotic description, the non-perturbative field theory sectors originate from spacetime-filling NS5-branes in an otherwise weakly coupled heterotic theory. In particular, the gravitational sector, and all states needed for the Distance Conjecture to hold, arise from the weakly coupled heterotic 
 sector. The appearance of the NS5-brane(s) in the heterotic frame, however, leaves its imprint in an exponential dependence of the prepotential in the heterotic dilaton of the form~\eqref{Fhet}. 

By Higgsing the strongly coupled field theory sector, such II$_0$ degenerations map to K-points in one-parameter Calabi--Yau moduli spaces.  We propose that in the heterotic picture, even after the Higgsing, the non-perturbative sector associated with the NS5-branes continues to leave its imprint in the exponential leading order term in the prepotential.
 The importance of this exponential dependence was observed and stressed in the recent analysis of K-points \cite{Hattab:2025aok}.
  Our main conclusion is that it does not contradict the fact that the gravitational sector is fully described by a weakly coupled heterotic string all the way to the heterotic string scale. In particular, both the Distance Conjecture and the Emergent String Conjecture hold as a consequence of the light towers identified in \cite{Friedrich:2025gvs, Monnee:2025ynn} associated with this perturbative heterotic string. Furthermore, both from the geometric understanding of the special Lagrangians associated with Gr$_2$ of \cite{Friedrich:2025gvs} and from the description of K-points in terms of a perturbative heterotic string theory in presence of a decoupling NS5-brane sector proposed in this note, we do not find evidence for the additional light states which, as pointed out in \cite{Hattab:2025aok}, could in principle underlie the effective field theory description of the K-point in the spirit of the Emergence Proposal.

Apart from what we believe is a reconciliation of  the K-point with the Emergent String Conjecture, our analysis leads to two interesting questions for future research. First, it would be desirable to better understand the transition between a  II$_0$ and a II$_1$ singularity directly in the dual heterotic frame. Our analysis suggests that this transition  should involve the nucleation of an NS5-brane along with a gauge bundle.
 While the details of the transition are irrelevant for the interpretation of the II$_0$ limits and hence of the K-point as such, it would be very interesting to understand it better from the perspective of the heterotic string, possibly along similar lines as in the 4d ${\cal N}=1$ transitions studied in \cite{Anderson:2022bpo}.

Second, 
the presence of an NS5-brane sector in heterotic string theory leads to a holomorphic anomaly for the elliptic genus of the heterotic string, as discussed in the context of the Emergent String Conjecture in \cite{Lee:2018spm, Lee:2020gvu, Lee:2020blx}. By Type IIA-heterotic duality, the elliptic genus can be interpreted as a generating function of certain relative Gopakumar--Vafa (GV) 
 invariants, which are encoded, by mirror symmetry to Type IIB theory, in the periods and hence the prepotential of the theory. It would be very interesting to see whether the characteristic exponential form of the prepotential, as in (\ref{Fhet}), can directly be traced back to the appearance of a holomorphic anomaly for the GV invariants that enter the heterotic elliptic genus. This would give further support for the picture developed in this note, and we hope to report on this question in the future. \\

{\bf Acknowledgements.}
We thank G. F. Casas, D. v.d. Heisteeg, S.-J. Lee, F. Marchesano, L. Melotti, M. Montero,  L. Paoloni, B. Pioline, T. Schimannek and especially J. Hattab and E. Palti for discussions and correspondence. TW thanks KIAS and Yonsei University, Seoul, for hospitality during completion of this work.
 This work is supported in part by Deutsche Forschungsgemeinschaft under Germany’s Excellence Strategy EXC 2121 Quantum Universe 390833306, by Deutsche Forschungsgemeinschaft through a German-Israeli Project Cooperation (DIP) grant “Holography and the Swampland” and by Deutsche Forschungsgemeinschaft through the Collaborative Research Center 1624 “Higher Structures, Moduli Spaces and Integrability.”

\appendix 
\section{Periods} \label{app-periods}

Here we include higher order terms in the expansion of the periods (\ref{eq:periods}), as determined with the help of \cite{Lee:2019oct} and \cite{Bastian:2021eom}:
\begin{widetext}

 \begin{eqnarray*}
   X^0 &=& \frac{1}{\pi} \left(1 + \beta^2 x_0 - \left( \frac{5}{216} - \frac{77 \beta^2}{108 } \right)x_0^2 + {\cal O}(x_0^2 x_1)\right)  \nonumber \\
    X^1 &=&  \frac{i}{\pi}\left( 
    1 -  \beta^2 x_0 - \left(\frac{5 }{216} + \frac{77  \beta^2}{108}\right) x_0^2 + {\cal O}(x_0^2 x_1)\right)\nonumber \\
 X^2 &=& -\frac{\beta}{\pi} \sqrt{x_0}\left(  1 + \frac{23}{54}  x_0 - \frac{1}{16}x_1 + \frac{23}{288} x_0 x_1 + {\cal O}(x_1^2)\right) \nonumber\\
F_2 &=& 
-\frac{i \, \beta \, \sqrt{x_{0}}}{\pi^{2}} \Bigg[
2 - \frac{1}{16} x_{1} - \frac{47}{2048} x_{1}^{2} 
 + \log(x_{1}) \left(
    \frac{1}{2} + \frac{23}{108} x_{0} - \frac{1}{32} x_{1}
    + \frac{23}{576} x_{0} x_{1} - \frac{15}{2048} x_{1}^{2}
\right)
\Bigg]\nonumber \\
F_1 &=& 
\frac{1}{\pi^{2}} \Bigg[
\frac{1}{2} \beta^{2} x_{0}
+ \left(- \frac{4}{81}
+ \frac{23}{162} \beta^{2} \right)x_{0}^{2} + {\cal O}(x_0^2 x_1)
 + \left(
    \frac{1}{8} - \frac{1}{8} \beta^{2} x_{0}
   +\left( - \frac{5}{1728}  - \frac{77}{864} \beta^{2}\right) x_{0}^{2} +  {\cal O}(x_0^2 x_1)
\right) 
 \log(x_0^2 x_{1})
\Bigg] \nonumber\\
F_0 &=& 
\frac{i}{\pi^{2}} \Bigg[
\frac{1}{2} \beta^{2} x_{0}
+ \left(\frac{4}{81} 
+ \frac{23}{162} \beta^{2}\right) x_{0}^{2} + {\cal O}(x_0^2 x_1)
 + \left(
    -\frac{1}{8} - \frac{1}{8} \beta^{2} x_{0}
    + \left(\frac{5}{1728} - \frac{77}{864} \beta^{2} \right)x_{0}^{2} + {\cal O}(x_0^2 x_1)
\right) 
 \log(x^2_{0}x_{1})
 \Bigg]
\end{eqnarray*}
 
 \end{widetext}

\bibliography{papers_Max}

\end{document}